\documentclass[dvips]{article}  
\usepackage{icrc2011}
\usepackage{lineno}

\usepackage{graphicx}  
\usepackage{multirow}
\usepackage{amssymb}

\newcommand{\etal}{\MakeLowercase{\textit{et al. }}} 

 \title{Online Gamma-Ray Burst catalog for neutrino telescopes.}
\shorttitle{J. A. Aguilar. GRB Online Catalog}

\authors{Juan A. Aguilar$^{1}$}
\afiliations{$^{1}$ Dept. of Physics, University of Wisconsin, 
           Madison, WI 53706, USA}
           
\email{aguilar@icecube.wisc.edu}
     
\abstract{The origin of cosmic rays is still one of the unresolved questions in modern physics. Violent and high energetic explosions of $\gamma$-ray emission known as Gamma Ray Bursts (GRBs) are perhaps one of the main candidates of sources of hadron acceleration and therefore of neutrino emission.
Neutrino telescopes search for signatures of cosmic neutrinos such as an excess of neutrinos in time and space coincidence with a GRB. These searches use catalogues of GRBs by satellite experiments. The online catalog presented here is a useful tool that provides a reference catalog of GRBs for neutrino telescopes in particular and GRBs analyzers in general.}


\begin{document}

\maketitle

\section{Introduction}

 Neutrino telescopes such as the IceCube Neutrino Observatory~\cite{IceCubeGRB} at the South Pole and ANTARES~\cite{AntaresGRB} in the Mediterranean Sea survey the entire sky using neutrinos as their cosmic messengers.  Since there is presently no identified source of high energy cosmic neutrinos these experiments usually perform a blind search all over the whole sky in order to identify an excess of neutrinos in any direction. However, it is possible to maximize the discovery potential and reduce the penalty from trials by pre-selecting a set of directions to look at based on possible neutrino candidate sources. Gamma Ray Bursts (GRBs) are among the best motivated neutrino emitters. GRBs are short flashes of $\gamma$-rays typically lasting from milliseconds to tens of seconds and emitting most of their energy in $>$~1 MeV photons. Their short duration makes them also very attractive since the time constraint enables a good rejection of the atmospheric neutrino and muon backgrounds making them background-free sources in neutrino astronomy. The online GRBs catalog {\bf grbweb}\footnote{http://grbweb.icecube.wisc.edu} is a dynamic list of GRBs based on the alert system of the Gamma-ray burst Coordinate Network (GCN)~\cite{GCN}. In the past other projects like GRBlog~\cite{GRBlog} had used the idea of parsing the information from the Gamma-ray burst Coordinate Network into a database but emphasizing the optical data from the GRB detection. The project presented here follows a similar approach but focusing mainly on the $\gamma$-ray detection information and providing neutrino spectrum information as part of the final output of the catalog. Several cross-references and validation checks are performed automatically to require minimal manual intervention although this may still be required due to the prose-style format of the GCN information. In Sec.~\ref{sec:parsing} the parsing of the GCN information and the database design are described. The description of the web interface as well as the different parts and features that compound it can be found in Sec.~\ref{sec:web}. The database management and the conflict control system are described in Sec.~\ref{sec:conflicts} and the conclusions and future plans for this online catalog are given in Sec.~\ref{sec:future}.

\section{Parsing and database design}
\label{sec:parsing}
 
As mentioned before, the information corresponding to the GRBs is extracted from the GCN system. This network distributes information about the location of Gamma-Ray bursts nearly in real-time in the form of the so-called {\it Notices}. These {\it Notices} are token-style messages with very limited information about GRBs. A more extensive report about the GRBs is given in the form of {\it {\it Circulars}}. The {\it {\it Circulars}} are e-mail messages with follow-up information about the GRB. These e-mails are sent to a central location and subsequent to an e-mail distribution list. The {\it Circulars} can also be obtained from the archive located in the GCN server. Unlike the {\it Notices}, {\it Circulars} provide detailed information about different GRBs parameters but in an unformatted way which makes the automatic parsing of the {\it Circulars} a challenging task.

In order to retrieve the information from the GCN {\it Circulars} and build the {\bf grbweb} online catalog, a set of PHP\footnote{http://www.php.net} scripts run every night downloading the new {\it Circulars} from the GCN archive and extracting the information provided by the GRB detector satellites. The values extracted for every interesting parameter are then inserted into a database hosted by a MySQL\footnote{http://www.mysql.com} server. Depending on the detector issuing the {\it Circular} different variables are provided. In addition to that, the writing style for each {\it Circular} differs depending on the detector and therefore a dedicated PHP script is used for each satellite and the information is inserted in individual database tables.  A description of the main tables in the database is given in Tab.~\ref{tab: database}.

\begin{table*}
\begin{center}
\begin{tabular}{ l |  c  | c | c| c| c| c| c| c| c |c| c| c |c|c} \hline
\multirow{2}{*}{Table title} & r.a.&dec.&err.&t90&t1&t2&\multirow{2}{*}{$\alpha$}&\multirow{2}{*}{$\beta$F}&E$_{peak}$&\multirow{2}{*}{$\gamma$}& E$_{min}$& E$_{max}$&$\mathcal{F}$ & \multirow{2}{*}{z} \\ 
 &\small{ J2000 ($^{\circ}$)} & \small{J2000 ($^{\circ}$)} & \small{($^{\circ}$)} & \small (s) & \small{(s)} & \small{(s)} &  &   & \small{ (keV) } &  &\small{(keV)}&\small{(keV)}& \small{(erg s$^{-1}$)} \\
\hline
Fermi LAT & $\checkmark$ & $\checkmark$& $\checkmark$& & & & & & & & & & &\\ \hline
Fermi GMB & $\checkmark$ & $\checkmark$  & $\checkmark$ &$\checkmark$ &$\checkmark$ & $\checkmark$&$\checkmark$ &$\checkmark$ & $\checkmark$ & &$\checkmark$ &$\checkmark$ &$\checkmark$ &\\  \hline
{\it Swift}-XRT& $\checkmark$ & $\checkmark$& $\checkmark$&$\checkmark$ & & & & & & & & & &\\  \hline
{\it Swift}-BAT & $\checkmark$ & $\checkmark$  & $\checkmark$ &$\checkmark$ &$\checkmark$ & $\checkmark$&$\checkmark$ &$\checkmark$ & $\checkmark$ &$\checkmark$ & & &\\  \hline
{\it Swift}/UVOT &$\checkmark$ & $\checkmark$& $\checkmark$& & & & & & & & & & &\\  \hline
Konus-Wind & &&&$\checkmark$ & $\checkmark$& $\checkmark$&$\checkmark$&$\checkmark$&$\checkmark$&&$\checkmark$&$\checkmark$&$\checkmark$&\\  \hline
Super-Agile& $\checkmark$ & $\checkmark$& $\checkmark$&$\checkmark$ &$\checkmark$ & & & & & & & & &\\ \hline
Suzaku WAM& & & &$\checkmark$ &$\checkmark$ &$\checkmark$ & $\checkmark$& $\checkmark$& $\checkmark$&$\checkmark$ &$\checkmark$ &$\checkmark$ & $\checkmark$&\\ \hline
Integral& $\checkmark$ & $\checkmark$& $\checkmark$& & $\checkmark$& & & & & & & &$\checkmark$ &\\ \hline
Redshift& & & & && & & & & & & & & $\checkmark$\\ 
\hline
\end{tabular}
\caption{\label{tab: database} The parameters r.a. and dec. are the right ascension and declination of the GRB; {\it err} is the position uncertainty, {\it t90} is the GRB reported duration; {\it t1} and {\it t2} are the start and stop time according to the satellite; $\alpha$, $\beta$, E$_{peak}$ are the first and second indexes and energy cutoff of the fitted Band function~\cite{Band} if such fit is provided, $\gamma$ is the spectral index in case of a single power-law fit of the $\gamma$-ray spectrum; $\mathcal{F}$ is the fluence and E$_{min}$, and E$_{max}$ are the extremes of the energy range in which the fluence is measured.}
\end{center}
\end{table*}

The database follows a relational structure, which means that all tables are related through a primary key used to connect all the common information of a given GRB without compromising the database integrity. An additional general table that keeps record of every {\it Circular} uploaded into the database is used to retrieve the information relating all entries of the same GRB from the different database tables. 

\section{The web interface}
\label{sec:web}

The final goal of the database is to provide a fast and simple way to obtain the main parameters characterizing a GRB and its spectrum. In order to retrieve that information a PHP web interface has been created: the web site {\bf grbweb} hosted in one of the IceCube web servers (http://grbweb.icecube.wisc.edu). On the main page the user can select one day or a period of time since the beginning of 2007. The PHP interface will connect to the database and create an on-the-fly catalog with all the main parameters for all the GRBs during the time interval selected. This catalog provides the basic information of each GRB and its expected neutrino spectrum. A detailed description of each of the variables displayed in the catalog is given in Tab.~\ref{tab: summary}. 

\begin{table}
\centering
\begin{tabular}{ l | l | c  } \hline
Section & Variable & Description \\
\hline
GRB &Name & GRB name in YYMMDD\\
 & UT & GRB time [UT]\\ \hline
{Positioning}&RA & Right ascension [$^{\circ}$] (J2000)\\
&Decl & Declination  [$^{\circ}$] (J2000)\\
&ERR & Uncertainty [$^{\circ}$]\\ \hline
{Timing}&T100& Duration defined as T2 - T1 [s]\\
&T1 & Start time [s]\\
&T2 & Stop time [s] \\ \hline
{$\gamma$-spectrum}&$\alpha_{\gamma}$ & First spectral index\\
&$\beta_{\gamma}$ & Second spectral index \\
&$\epsilon_{\gamma}$ & Energy peak [keV]\\
&$\mathcal{F}_{\gamma}$ & Fluence [erg cm$^{-2}$]\\
&E$_{min}$ & Minimum energy [keV]\\
&E$_{max}$ & Maximum energy [keV]\\ \hline
{Other}& z & redshift \\ \hline
{$\nu$-spectrum}&$f_{\nu}$ & Flux at $\epsilon_{1}$ [GeV$^{-1}$ cm$^{-2}$]\\
&$\epsilon_{1}$ & First energy break [PeV]\\
&$\epsilon_{2}$ & Second energy break [PeV]\\
&$\alpha_{\nu}$ & WB index before $\epsilon_{1}$\\
&$\beta_{\nu}$ & WB index between $\epsilon_{1}$ and $\epsilon_{2}$\\
&$\gamma_{\nu}$ & WB spectrum after $\epsilon_{2}$\\
\hline
\end{tabular}
\caption{\label{tab: summary}
List and description of the parameters on the online catalog.}
\end{table}

Different  satellites may report different values for the same parameter. As an example the GRB coordinate position depends on the detector pointing resolution and therefore different values are provided depending on the {\it Circular}. In that case, the web interface takes a decision based on the instrument resolution or accuracy in measuring that specific parameter. The exact ordering is described in Tab.~\ref{tab:order} for three parameter classifications: timing information, positioning, and photon spectrum parameters. 

\begin{table}
\centering
\begin{tabular}{ c |  c|  c | c } 
\hline
Order & Timing & Positioning & $\gamma$-spectrum  \\
\hline
1 & Fermi GBM & {\it Swift}/UVOT & Fermi GBM \\
2 & Konus-Wind & {\it Swift}/XRT & Konus-Wind \\
3 & Suzaku WAM& {\it Swift}/BAT &  Suzaku WAM\\
4 & {\it Swift}/BAT & SuperAgile & {\it Swift}/BAT \\
5 & Integral & Fermi LAT &  \\
6 & Suzaku WAM & Fermi GBM& \\
7 &  & Integral &  \\
\hline
\end{tabular}
\caption{\label{tab:order}
Ordering of parameter values depending on the {\it Circular}.}
\end{table}

\begin{figure}[t]  
\centering
\includegraphics[width=3.in]{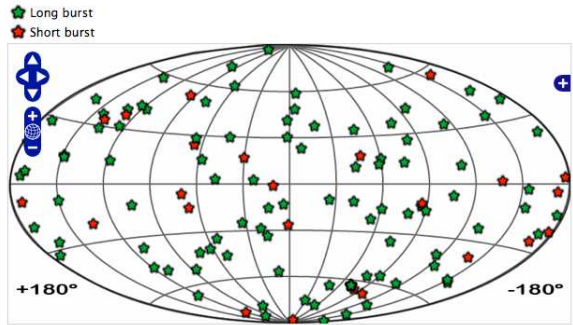}
\caption{\label{fig: skymap} Sky map of all GRBs from 1$^{st}$ January until 31$^{st}$ December 2010 in galactic coordinates. The red stars indicates short ($<$2~s) GRBs while green stars are long ($\ge$2~s) GRBs.}
\end{figure}

The catalog table is a dynamic html table in which the user can apply further filters as well as download the selected GRB list into a text file. 
Another feature allows the user to automatically plot the coordinates of all selected GRB on a skymap. Figure~\ref{fig: skymap} shows the skymap of all GRB corresponding to the year 2010 automatically generated by the {\bf grbweb}. The open code OpenLayers\footnote{http://openlayers.org/} is used to display the map. By clicking on the GRB coordinates icons a pop-up window will appear showing some basic information for the specific GRB.

\subsection{Neutrino spectrum}

As can be seen from Tab.~\ref{tab: summary}, together with the parsed information from the GCN {\it Circulars} the {\bf grbweb} interface will also calculate the predicted neutrino spectrum according to the Waxman-Bahcal model~\cite{WB}. The calculations followed on the site are described in detail in Guetta et al. (\cite{Guetta}). This theory assumes that neutrinos come from p-$\gamma$ interactions near the break energy of the GRB, making E$_{peak}$ the second most important parameter, after fluence, for estimating the neutrino flux. The web site uses the $\gamma$-spectrum variables to calculate the neutrino spectrum for each GRB. In case of missing or incomplete information, default values are used. These default values are selected depending on the length of the $\gamma$-ray emission of the GRB. The GCN distinguishes between short-hard GRBs ($<$2~s) and long-soft GRBs ($\geq$2~s) that might have different underlying source. Table~\ref{tab: wb} summarizes these default values used for the neutrino spectrum calculation. As in the case of the GRBs skymap, the user can select a list of GRBs and plot the corresponding neutrino spectra in an automatically generated plot. The neutrino spectra plot is generated using jpGraph\footnote{http://jpgraph.net/}, a PHP graph library for non-comercial use. 


\begin{table}[hc]
\centering
\begin{tabular}{ l |  c |c } 
\hline
Variable & Long Soft Burst &  Short Hard Burst \\
\hline
Fluence & $10^{-5}$ erg cm$^{-2}$ & $10^{-5}$ erg cm$^{-2}$\\
z & 2.15 & 0.5 \\
E$_{peak}$ & 200 keV & 1000 keV \\
$\alpha$ & 1 & 1 \\
$\beta$ & 2 & 2 \\
$L_{iso}$ & $10^{52}$ erg & $10^{51}$ erg \\
$\Gamma_{jet}$ & 316 & 316 \\
t$_{var}$ & 0.01 & 0.001\\
$\epsilon_{e}$ & 0.1 & 0.1\\
$\epsilon_{B}$ & 0.1 & 0.1\\
f$_{e}$ & 0.1 & 0.1\\
\hline
\end{tabular}
\caption{\label{tab: wb}
Average Waxman-Bahcal values used as default values for the neutrino spectrum calculation.}
\end{table}

\subsection{The GRB view}

The online catalog gives the basic information for all GRBs in the selected time interval. However it is possible to access more detailed information for an specific GRBs by clicking the link on the GRB name. This will open the GRB view, a window that shows all the collected information stored in the database for a particular GRB. From this window the user can see which satellites observed the GRBs and the information provided. The GRB view provides direct links to the GCN {\it Circulars} used to extract the information. This way the user can directly compare the parsed information in the database with respect to the actual {\it Circulars}. 


\subsection{Light Curve view}

\begin{figure}[ht]  
\centering
\includegraphics[width=3.in]{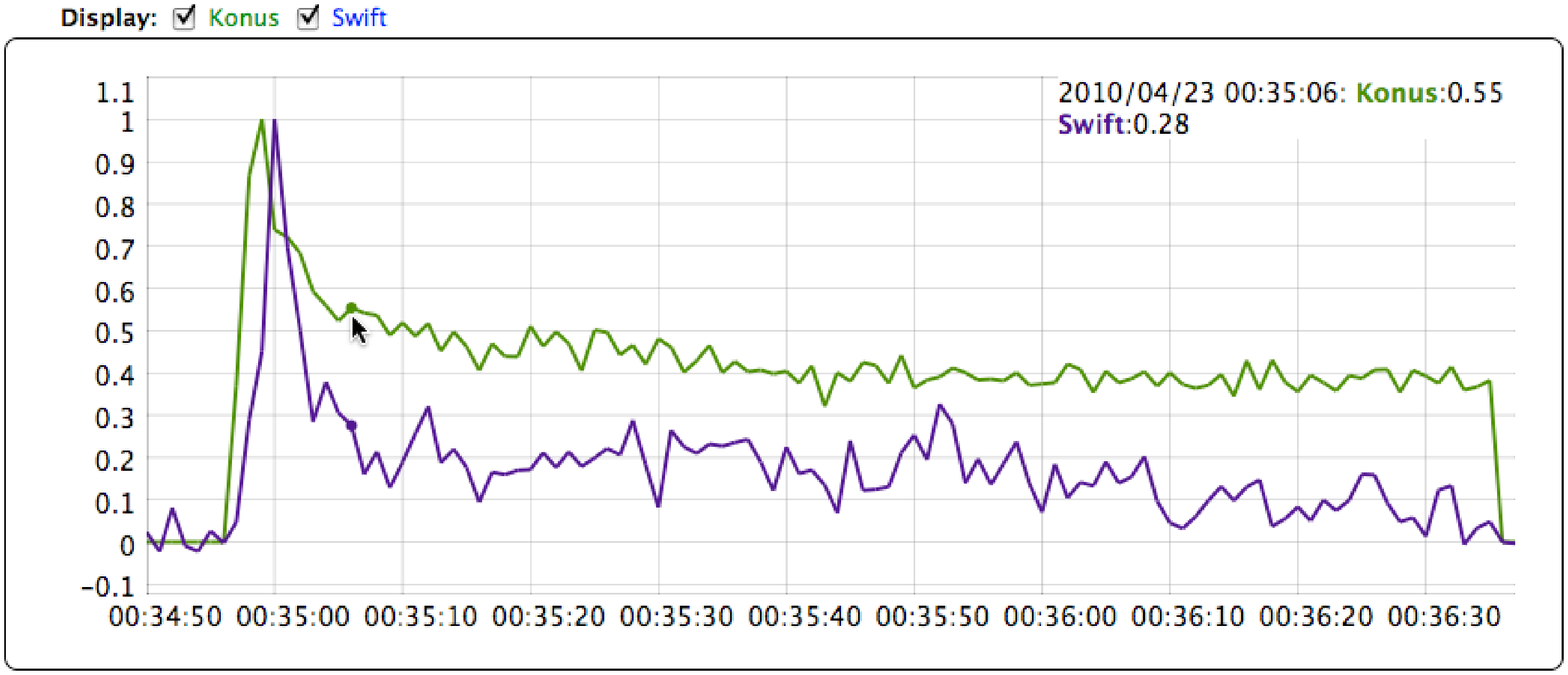}
\caption{\label{fig: LightCurve} Konus and Swifht light curves provided for GRB100423A. The light curves have been normalize to its maximum.}
\end{figure}

If the photon light curve information has been measured and is publicly available the PHP scripts will also parse and upload that information into the database. In these cases the online catalog will also provide a link to the light curve for the GRB. This link will open the Light Curve view window (see Fig.~\ref{fig: LightCurve}) where different light curves (as many as they have been measured) are displayed. Currently up to three satellites are searched for light curve measurements: Integral, {\it Swift}, and Konus-Wind.  Figure~\ref{fig: LightCurve} shows the light curves for GRB100423 measured by Konus-Wind and {\it Swift}. This view is also an interactive graph in which the user can select different time ranges.

\section{Conflicts and supervision}
\label{sec:conflicts}

An important goal of the {\bf grbweb} is to be able to run with a minimal human supervision. The prose-style of the {\it Circulars} however makes this a challenging task and therefore as part of the system different validation checks are performed when parsing the GRB information in order to verify the integrity of the information uploaded to the database. Occasionally satellites send more than one {\it Circular} for a single GRB generally correcting or improving some of the information provided in the previous one. This duplication of {\it Circulars} from the same satellite can lead to inconsistencies in the catalog. For that reason when two different values are reported for the same parameter the catalog will display an alert icon next to the GRB name and the affected parameter value will be displayed in red on the catalog list. 
In some other cases typos and misspelling in the {\it Circulars} can lead to parsing errors. When creating the catalog some consistency checks are performed in order to validate the values provided by different satellites. As an example if the GRB coordinates given by one 
satellite differ significantly from the coordinates from another detector the catalog will also display the alert icon. 
The presence of the alert symbol does not imply that the information provided for that GRB is incorrect but suggests the user carefully compare the values with the original {\it Circulars}. These conflicts can be resolved by the administrator by manually modifying the database when necessary. The goal of the conflict system is to reduce significantly the amount of time of human supervision of the catalog, ensuring in this way an almost autonomous maintenance of the database.

\section{Conclusion and future plans}
\label{sec:future}

The web site {\bf grbweb.icecube.wisc.edu} is now fully operational and publicly accessible. New {\it Circulars} are parsed and incorporated into the database on a daily bases. The parsing has been demonstrated to be stable despite the un-formatted style of the {\it Circulars} and the conflict system helps significantly in management of the database. An machine-readable format of the GCN {\it Circulars} would significantly help this and similar projects in the future. New features for the online catalog can be added as, for example, other neutrino spectra calculations besides the Waxman-Bahcall model. New light curves from other satellites will be incorporated into the database, as well as the parsing of {\it Circulars} other than those emitted by $\gamma$-ray detectors. For example, optical and x-ray {\it Circulars} providing this way a more general view of all the parameters measured for the GRB would extend the scope of this online catalog to the optical and x-ray astronomy.

\section{Acknowledgements}
The author would like to thank Kyle Jero for his valuable help during the beginning of this project. Also the whole IceCube collaboration for their support and the GRB working group for the feedback.

\end{document}